\documentclass[aps,twocolumn,pra,showpacs,superscriptaddress]{revtex4}
\usepackage{amstext}
\usepackage{latexsym}
\usepackage{bm}
\usepackage{amssymb}
\usepackage{amsmath}
\usepackage{graphicx}
\usepackage{subfigure}
\usepackage{natbib}
\usepackage{epsfig}
\usepackage{amsfonts}
\usepackage{mathrsfs}
\usepackage{CJK}
\usepackage[toc,page,title,titletoc,header]{appendix}
\begin{document}

\title{The scattering amplitude of ultracold atoms near the p-wave magnetic
Feshbach Resonance}
\author{Peng Zhang}
\affiliation{ERATO, JST, Macroscopic Quantum Control Project, Hongo,
Bunkyo-Ku, Tokyo 113-8656, Janan }
\affiliation{Department of
Physics, Renmin University of China, Beijing, 100190, China}

\author{Pascal Naidon}
\affiliation{ERATO, JST, Macroscopic Quantum Control Project, Hongo,
Bunkyo-Ku, Tokyo 113-8656, Janan }

\author{Masahito Ueda}
\affiliation{ERATO, JST, Macroscopic Quantum Control Project, Hongo,
Bunkyo-Ku, Tokyo 113-8656, Janan }
 \affiliation{Department of
Physics, University of Tokyo, Hongo, Bunkyo-Ku, Tokyo 113-8656,
Janan }

\begin{abstract}
Most of the current theories on the p-wave superfluid in cold atomic gases
are based on the effective-range theory for the two-body scattering, where
the low energy p-wave scattering amplitude $f_1(k)$ is given by $%
f_1(k)=-1/[ik+1/(\mathcal{V}k^2)+1/\mathcal{R}]$, where $k$ is the
incident momentum, and $\mathcal{V}$ and $\mathcal{R}$ are the
$k$-independent scattering volume and effective-range, respectively.
However, due to the long-range nature of the van der Waals
interaction between two colliding ultracold atoms, the p-wave
scattering amplitude of the two atoms is not described by the
effective-range theory \cite{JMP1, gao1}. In this paper we provide
an explicit calculation for the p-wave scattering of two ultracold
atoms near the p-wave magnetic Feshbach resonance (PMFR). We show
that the low energy p-wave scattering amplitude in the presence of
PMFR takes the
form $f_1(k)=-1/[ik+1/(\mathcal{V}^{\mathrm{eff}}k^2)+1/(\mathcal{S}^{\mathrm{eff}%
}k)+1/\mathcal{R}^{\mathrm{eff}}]$ where $\mathcal{V}^{\mathrm{eff}},$ $%
\mathcal{S}^{\mathrm{eff}}$ and $\mathcal{R}^{\mathrm{eff}}$ are $k$%
-dependent parameters. Based on this result, we show sufficient conditions
for the effective range theory to be a good approximation of the exact
scattering amplitude. Using these conditions we show that the
effective-range theory is a good approximation for the p-wave scattering in
the ultracold gases of $^{6}$Li and $^{40}$K when the scattering volume is
enhanced by the resonance.
\end{abstract}

\maketitle

\section{Introduction}

Recently, ultracold atomic gases with strong p-wave interaction have
attracted broad interest both experimentally \cite%
{expK40, expK40b, expK40Mola, expK40Dipo, expK40Molb, expLi6a,
expLi6b, expLi6c, expLi6MolSca, expLi6R87, expLiK, expLi6Mol,
expKRb}
and theoretically \cite%
{pYou,2DpSFa,pSFi,2DpSFb,pSF1cERTa,pSF1cERTb,pSFERTa,pSFa,p2DSFb,pSFf,2DpSFc,pSFh,p2DSFa,pSFc,
1DpSFb,pSFe,psSFa,pSFERTc,GPRA,
1DpSFa,pSFb,pSFERTb,pSF1cERTc,pFRvortex,pSFOLa,pBosonERT,pSFBoseFermi,twoFermionERTa,MaqidaJPSJ1,
MaqidaJPSJ2,ZinnerEPJD,MaqidaPRL1,PRA2010a,MaqidaPRA1,MaqidaPRA2,MaqidaPRA3,XXXa,StoofPRL,
p3bERT,pPseudoPotentialERTa,p2bFR,
pPseudoPotentialERTb,pPseudoPotentialc,pPseudoPotentiald,pPseudoPotentiale,PRLBlume,nf1,nf2,nf3,nf4}%
. The p-wave magnetic Feshbach resonances, which can generate tunable p-wave
ineratomic interactions, have been observed in the cold gases of $^{\mathrm{%
40}}$K \cite{expK40,expK40b,expK40Dipo}, $^{\mathrm{6}}$Li \cite%
{expLi6a,expLi6b}, $^{\mathrm{6}}$Li-$^{\mathrm{87}}$Rb mixture \cite%
{expLi6R87}, $^{\mathrm{6}}$Li-$^{\mathrm{40}}$K mixture \cite{expLiK}, and $%
^{\mathrm{40}}$K-$^{\mathrm{87}}$Rb mixture \cite{expKRb}. The p-wave
Feshbach molecules have also been created and studied in the gases of $^{%
\mathrm{40}}$K \cite{expK40Mola,expK40Molb} and $^{\mathrm{6}}$Li \cite%
{expLi6a,expLi6c,expLi6Mol,expLi6MolSca}. These experimental
achievements stimulate theoretical researches on the quantum
superfluid in ultracold atomic gases
with strong p-wave interactions \cite%
{pYou,2DpSFa,pSFi,2DpSFb,pSF1cERTa,pSF1cERTb,pSFERTa,pSFa,p2DSFb,pSFf,2DpSFc,pSFh,p2DSFa,pSFc,
1DpSFb,pSFe,psSFa,pSFERTc,GPRA,
1DpSFa,pSFb,pSFERTb,pSF1cERTc,pFRvortex,pSFOLa,pBosonERT,pSFBoseFermi,MaqidaJPSJ1,MaqidaJPSJ2,
ZinnerEPJD,MaqidaPRL1,PRA2010a,MaqidaPRA1,MaqidaPRA2,MaqidaPRA3,XXXa,StoofPRL}%
, as well as the relevant few-body problems
\cite{twoFermionERTa,StoofPRL,
p3bERT,pPseudoPotentialERTa,p2bFR,pPseudoPotentialERTb,pPseudoPotentialc,pPseudoPotentiald,pPseudoPotentiale,PRLBlume,nf1,nf2,nf3,nf4}%
.

Until now, most theories of ultracold atomic gases with strong p-wave
interactions \cite%
{pSF1cERTa,pSF1cERTb,pSFERTa,pSFERTc,pSFERTb,pSF1cERTc,pBosonERT,StoofPRL,twoFermionERTa,p3bERT,pPseudoPotentialERTa,p2bFR,pPseudoPotentialERTb}
are based on the low energy expansion of the p-wave scattering amplitude $%
f_{1}(k)$ given by the effective-range theory \cite{taylor}
\begin{eqnarray}
f_{1}(k)=-\frac{1}{ik+\frac{1}{\mathcal{V}k^{2}}+\frac{1}{\mathcal{R}}}.
\label{1a}
\end{eqnarray}%
Here $\vec{k}$ is the relative momentum of the two atoms; $\mathcal{V} $ is
the scattering volume and $\mathcal{R}$ is the effective-range. The
effective-range theory for the scattering amplitude is used in both theories
of p-wave atomic superfluids \cite%
{pSF1cERTa,pSF1cERTb,pSFERTa,pSFERTc,pSFERTb,pSF1cERTc,pBosonERT,StoofPRL}
and related few-body problems \cite{twoFermionERTa,StoofPRL,
p3bERT,pPseudoPotentialERTa,p2bFR,pPseudoPotentialERTb}. In particular, the
separable two-body potential \cite{pSFERTa} $V(k,k^{\prime
})=\lambda_1w(k)w(k^{\prime })$ used in the many-body Hamiltonian for the
p-wave atomic superfluid is derived directly from low energy expansion (\ref%
{1a}) of the scattering amplitude.

However, the effective-range theory is correct only for the
\textit{short range} potentials (\textit{e.g.}, Yukawa potential)
\cite{taylor} which decays faster than any power function
$r^{-\gamma }$ in the large interatomic distance limit $r\rightarrow
\infty $. Here $\vec{r}$ is the relative coordinate between two
atoms. For a realistic interaction between two cold atoms, which is
described by a \textit{long-range} potential dominated by the van
der Waals term $-\hbar^2\beta_{6}^4/(r^{6}m)$ in the limit
$r\rightarrow \infty $. Here $m$ is the single-atom mass, $\hbar$ is
the Plank constant and $\beta_6$ is the van der Waals length. Due to
the long-range van der Waals potential, the effective-range theory
and the low-energy expansion (\ref{1a}) of the p-wave scattering
amplitude is not applicable any longer \cite{JMP1,gao1}.

In the presence of a p-wave magnetic Feshbach resonance (PMFR) in the
ultracold gases of polarized fermionic atoms, the p-wave scattering
amplitude of the atomic collision is contributed by both the background
potential in the open channel and the bound state in the closed channel. The
long-range nature of the background potential makes the final scattering
amplitude to be inconsistent with the effective-range theory.

Therefore, it is essential to investigate the condition under which the
effective-range theory (\ref{1a}) can be used as an approximation of the
exact p-wave scattering amplitude under a PMFR. If the effective-range
theory provides a good approximation of the scattering amplitude, then the
previous theories on p-wave superfluid would be applicable; nevertheless, if
the exact scattering amplitude is found to be significantly different with
the one in Eq. (\ref{1a}), then the previous theories should be modified.
Especially, the separable two-body potential cannot be used any more.

The low-energy p-wave scattering amplitude near PMFR has been investigated
in Refs. \cite{p2bFR,p3bERT}. However, these studies are based on simplified
models of the atomic interaction, e.g., zero background potential \cite%
{p2bFR} or a separable background potential that decays exponentially in the
momentum space \cite{p3bERT}. The long-range van der Waals potential is not
taken into account in either case. Due to these simplifications, the
scattering amplitudes given in Refs. \cite{p2bFR,p3bERT} automatically have
the form of Eq. (\ref{1a}), and cannot be used to judge the applicability of
the effective-range theory.

In this paper, based on the realistic long-range inter-atomic potential, we
provide an explicit calculation for the low-energy p-wave scattering
amplitude of two spin polarized fermonic atoms near a PMFR, and then discuss
the condition under which the effective-range theory can be used as a good
approximation. We get sufficient conditions for the effective-range theory,
and show that for the ultracold gases of of $^{6}$Li and $^{40}$K with the
Fermi temperature of the order 1$\mu$K, the effective-range theory can be
used as a good approximation in the resonance regime where the scattering
volume is enhanced

\subsection{Main results}

The main results of this paper are summarized as follows.

In this work we first calculate the exact expression of the low energy
p-wave scattering amplitude with a PMFR. We prove that the scattering
amplitude can be expressed as
\begin{eqnarray}
&&f_{1m_{z}}(k)  \nonumber \\
&=&-\frac{1}{ik+\frac{1}{\mathcal{V}^{\mathrm{eff}}(k;B;m_{z})k^{2}}+\frac{1%
}{\mathcal{S}^{\mathrm{eff}}(k;B;m_{z})k}+\frac{1}{\mathcal{R}^{\mathrm{eff}%
}(k;B;m_{z})}}.  \nonumber
\\&&  \label{1c}
\end{eqnarray}%
with $B$ the strength of the magnetic field applied along the $z$ axis. Here
$m_z$ is the $z$-component of the angular momentum; $\mathcal{V}^{\mathrm{eff%
}},$ $\mathcal{S}^{\mathrm{eff}}$ and $\mathcal{R}^{\mathrm{eff}}$ are $k$%
-dependent scattering parameters. We obtain the general expressions for
them. It is pointed out that, the denominator of $f_{1m_{z}}(k)$ cannot be
expressed as a Laurent series with $k$-independent coefficients because $%
\mathcal{V}^{\mathrm{eff}},$ $\mathcal{S}^{\mathrm{eff}}$ and $\mathcal{R}^{%
\mathrm{eff}}$ are not analytical functions of $k$.

Equation (\ref{1c}) shows that, in the presence of a PMFR the inconsistency
of the scattering amplitude with the effective-range theory is displayed in
a more complicated manner. The low-energy p-wave scattering amplitude in Eq.
(\ref{1c}) is different from the one obtained from the standard
effective-range theory in the following two senses:

1. The scattering parameters $(\mathcal{V}^{\mathrm{eff}},\mathcal{S}^{%
\mathrm{eff}},\mathcal{R}^{\mathrm{eff}})$ depend on the incident momentum $%
k $.

2. The term $1/[\mathcal{S}^{\mathrm{eff}}(k;B;m_{z})k]$ cannot be included
in the effective-range theory.

After obtaining the p-wave scattering amplitude under a PMFR, we
discuss the applicability of the effective rang theory as an
approximation of the scattering amplitude (\ref{1c}). We find that,
in the BEC side of the PMFR where $\mathcal{V}^{\mathrm{eff}}$ and
$\mathcal{R}^{\mathrm{eff}}$ have the same sign, sufficient
conditions for the validity of effective-range theory are
$r_{1},r_{2}<<1$. In the BCS side of the resonance, the sufficient
conditions become $r_{1},r_{2},r_{3}<<1$. Here $r_{1}$, $r_{2}$ and
$r_{3}$ are defined in Eqs. (\ref{r1new}), (\ref{r2}) and
(\ref{r3}). If these conditions are satisfied, the scattering
amplitude of our system can be approximated as
\begin{eqnarray}
f_{1m_{z}}(k)\approx
-\frac{1}{ik+\frac{1}{\mathcal{V}^{\mathrm{eff}}(0;B;m_{z})k^{2}}+\frac{1}{\mathcal{R}^{\mathrm{eff}}(0;B;m_{z})}}\nonumber\\
 \label{1d}
\end{eqnarray}
which has the same form as Eq. (\ref{1a}) derived by the
effective-range theory.

Qualitatively speaking, the above sufficient conditions means that
we can use the effective-range theory if the fermonic momentum of
the cold gas is low enough, the magnetic field is tuned close enough
to the resonance point and the background scattering potential in
the open channel is far away from the zero-energy shape resonance
point. For the realistic cold gases of Fermi atoms, if the
background scattering is far away from the shape resonance, the
effective range theory can usually be used in the total region where
the $p$-wave interaction is negligible.

The paper is organized as follows. In Sec. II, we calculate the p-wave
scattering amplitude near a PMFR, and obtain the low energy expansion in (%
\ref{1c}). The parameters $(\mathcal{V}^{\mathrm{eff}},\mathcal{S}^{\mathrm{%
eff}},\mathcal{R}^{\mathrm{eff}})$ are expressed in terms of the background
scattering parameters and the magnetic field. In Sec. III we discuss the
range of applicability of the effective-range theory. We show that, under
some simple sufficient conditions $r_1,r_2<<1$ or $r_1,r_2,r_3<<1$, the
scattering amplitude given by the effective-range theory is a good
approximation for the exact one obtained in Sec. II. We further show that
these conditions are well satisfied in the cold gases of $^{\mathrm{40}}$K
and $^{\mathrm{6}}$Li when the scattering volume is enhanced by a PMFR, and
then the previous results based on the effective-range theory are applicable
for these systems. In Sec. IV there are some conclusion and discussions. We
describe some detail of our calculations in the appendixes.

\section{Low-energy scattering amplitude near the p-wave Feshbach resonance}

\subsection{p-wave phase shifts with PMFR}

In this section we calculate the p-wave scattering amplitude in presence of
a PMFR induced by a magnetic field along the $z$ direction. We begin with
the two-channel Hamiltonian for the relative motion of two atoms (Fig. 1):
\begin{eqnarray}
H=\left(
\begin{array}{cc}
\hat{T}+V^{(\mathrm{{{bg})}}}(r) & W(r) \\
W(r) & \hat{T}+V^{(\mathrm{cl})}(\vec{r})+\varepsilon (B)%
\end{array}%
\right),  \label{hamiltonian}
\end{eqnarray}%
where $\hat{T}$ is the kinetic energy of relative motion, $V^{(\mathrm{{{bg})%
}}}(r)$ is the background scattering potential in the open channel, $W(r)$
is the coupling between the open and closed channel, and $V^{(\mathrm{cl})}(%
\vec{r})$ is the interaction potential in the close channel, that has a $B$%
-dependent positive threshold $\varepsilon (B)$. In this paper, for
simplicity, we assume that the background potential $V^{\mathrm{bg}}(r)$ is
independent on the direction of $\vec{r}$ and invariant under the SO(3)
rotation. We further assume that, in the closed channel there are only three
bare p-wave bound states $|\phi _{\mathrm{res}}^{(m_{z})}\rangle $ which are
near resonance with the threshold of the open channel. Here $m_{z}=0,\pm 1$
is the angular momentum along the $z$ axis. The self energy $E_{m_{z}}^{(%
\mathrm{{{cl})}}}(B)=\mu _{\mathrm{res}}(B-B_{\mathrm{res}}^{(m_{z})})$ of $%
|\phi _{\mathrm{res}}^{(m_{z})}\rangle $ is determined by the strength of
the magnetic field. The difference between $B_{\mathrm{res}}^{(m_{z})}$ with
$m_{z}=0,\pm 1$ depends on the atomic magnetic dipole. For the atoms with
small magnetic dipole, \textit{e.g.,} $^{6}$Li, the values of $B_{\mathrm{res%
}}^{(0)}$ and $B_{\mathrm{res}}^{(\pm 1)}$ are close with each other \cite%
{p2bFR}, while for the the atoms with large magnetic dipole, \textit{e.g.,} $%
^{40}$K, the difference between $B_{\mathrm{res}}^{(0)}$ and $B_{\mathrm{res}%
}^{(\pm 1)}$ is quite large \cite{expK40Dipo}.

\begin{figure}[tbp]
\includegraphics[bb=73bp 313bp 578bp 587bp,clip,width=8cm]{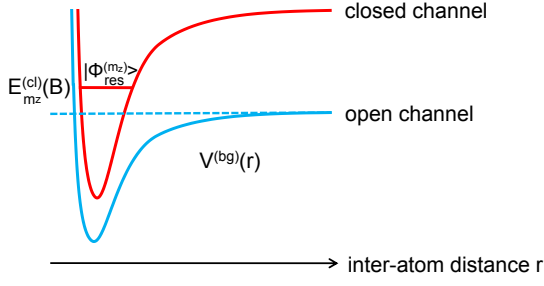}
\caption{(color online) The two channel model of the p-wave Feshbach
resonance.}
\end{figure}

The p-wave scattering amplitude in the open channel can be defined with the
standard scattering theory \cite{taylor}. To this end we firstly introduce
the two-component stationary scattering state%
\begin{eqnarray}
|\Phi _{\vec{k}}^{(+)}\rangle \equiv \left(
\begin{array}{c}
|\phi _{\vec{k}}^{(\mathrm{op}+)}\rangle \\
|\phi _{\vec{k}}^{(\mathrm{cl}+)}\rangle%
\end{array}%
\right) =\Omega _{+}\left(
\begin{array}{c}
|\vec{k}\rangle \\
0%
\end{array}%
\right)  \label{phik}
\end{eqnarray}%
where $|\vec{k}\rangle $ is the eigenstate of the atomic relative momentum
with eigen value $\hbar\vec{k}$, and $\Omega _{+}$ is the M\o {}ller
operator defined as%
\begin{eqnarray}
\Omega _{+}=\lim_{\tau \rightarrow \infty }e^{-iH\tau/\hbar}e^{i\hat{T}%
\tau/\hbar}.
\end{eqnarray}%
In the large interatomic distance limit $r\rightarrow \infty $, the
asymptotic behavior of the state $|\Phi _{\vec{k}}^{(+)}\rangle $ can be
expressed as%
\begin{eqnarray}
\langle \vec{r}|\Phi _{\vec{k}}^{(+)}\rangle =\frac{1}{(2\pi\hbar)^{\frac{3}{%
2}}}\left(
\begin{array}{c}
e^{i\vec{k}\cdot \vec{r}}+f(\hat{r},\vec{k})\frac{e^{ikr}}{r} \\
0%
\end{array}%
\right)  \label{asypsi}
\end{eqnarray}%
with $\hat{r}=\vec{r}/r$ and $f(\hat{r},\vec{k})$ the scattering amplitude
which can be further expanded in terms of different partial waves:%
\begin{eqnarray}
f(\hat{r},\vec{k})=4\pi \sum_{lm_{z}}f_{lm_{z}}(k)Y_{l}^{m_{z}}(\hat{r}%
)Y_{l}^{m_{z}}(\hat{k})^{\ast }.  \label{parf}
\end{eqnarray}%
Here $Y_{l}^{m_{z}}(\hat{r})$ is the spherical harmonic function. For a
scattering potential with SO(3) symmetry, the partial wave scattering
amplitude only depends on the quantum number of the angular momentum $l$. In
our case, the SO(3) symmetry is broken by the interaction between the atomic
magnetic dipole and the magnetic field. Then we have a $m_{z}$-dependent
scattering amplitude $f_{lm_{z}}(k)$.

In the case of low-energy scattering between two spin polarized fermonic
atoms, one can neglect all the high partial wave scattering amplitudes $%
f_{lm_{z}}(k)$ with $l\geqslant 2$, and only consider the p-wave amplitudes $%
f_{1m_{z}}(k)$, which can be further expressed in terms of the p-wave phase
shifts $\delta _{1m_{z}}(k)$:%
\begin{eqnarray}
f_{1m_{z}}(k)=-\frac{1}{ik-k\cot \delta _{1m_{z}}(k)}.  \label{fdelta}
\end{eqnarray}

During the scattering process, the bare bound state $|\phi _{\mathrm{res}%
}^{(m_{z})}\rangle $ is coupled with the p-wave background scattering states
in the open channel and significantly change the p-wave scattering amplitude
$f_{1m_{z}}(k)$. This effect can be directly treated via the Feshbach
resonance theory, \textit{e.g.}, the methods in Ref. \cite{fano} and Ref.
\cite{julian}.

After a straightforward calculation in Appendix A, we find that the final
phase shift $\delta _{1m_{z}}(k)$ is the sum of the background phase shift $%
\delta _{1}^{\mathrm{(bg)}}(k)$ for the background potential $V^{(\mathrm{{{%
bg})}}}(r)$\ and a correction $\Delta _{1m_{z}}(k)$ given by the closed
channel:
\begin{eqnarray}
\delta _{1}(k)=\delta _{1}^{\mathrm{(bg)}}(k)+\Delta
_{1m_{z}}(k).\label{delta1a}
\end{eqnarray}
Here $\Delta _{1m_{z}}(k)$ satisfies%
\begin{eqnarray}
-k{\cot }\Delta _{1m_{z}}(k)=\frac{k}{\pi }\frac{\hbar
^{2}k^{2}/m-E_{m_{z}}^{(\mathrm{{{cl})}}}(B)-g_{m_{z}}(k^{2})}{|\langle
\phi _{\mathrm{res}}^{(m_{z})}|W|\psi
_{k1m_{z}}^{(\mathrm{bg}+)}\rangle |^{2}}.\label{delta1}
\end{eqnarray}
with $|\psi _{k1m_{z}}^{(\mathrm{bg}+)}\rangle $ and $g_{m_{z}}(k^{2})$
given by%
\begin{eqnarray}
|\phi _{\vec{k}}^{(\mathrm{bg}+)}\rangle  &=&\left( \frac{2}{m\hbar k}%
\right) ^{\frac{1}{2}}\sum_{l,m_{z}}|\psi _{k1m_{z}}^{(\mathrm{bg}+)}\rangle
Y_{l}^{m_{z}}(\hat{k})^{\ast },  \label{psik1mz} \\
g_{m_{z}}(k^{2}) &=&\mathrm{Re}\langle \phi _{\mathrm{res}%
}^{(m_{z})}|WG_{+}^{(\mathrm{bg})}(k^{2})W|\phi _{\mathrm{res}%
}^{(m_{z})}\rangle \mathrm{.}
\end{eqnarray}%
In the above we have used the background Green's function $G_{+}^{(\mathrm{{%
bg})}}(k^{2})$
\begin{eqnarray}
G_{+}^{(\mathrm{{bg})}}(k^{2})=\frac{1}{\hbar ^{2}k^{2}/m+i0^{+}-\hat{T}-V^{(%
\mathrm{{{bg})}}}}.
\end{eqnarray}
and the background scattering state $|\phi _{\vec{k}}^{(\mathrm{bg}%
+)}\rangle $ defined as%
\begin{eqnarray}
|\phi _{\vec{k}}^{(\mathrm{bg}+)}\rangle =|\vec{k}\rangle +G_{+}^{(\mathrm{{%
bg})}}(k^{2})V^{(\mathrm{{{bg})}}}|\vec{k}\rangle .\label{psibg}
\end{eqnarray}

In the following subsections, we evaluate the low-energy expression of the
scattering amplitude $f_{1m_{z}}(k)$ by expanding the term $%
-k\cot\delta_{1m_{z}}(k)$ in Eq. (\ref{fdelta}) in the limit $k\rightarrow 0$%
. As shown in Eq. (\ref{delta1a}), the phase shift $\delta_{1m_{z}}(k)$ is
the sum of $\delta _{1}^{\mathrm{(bg)}}(k)$ and $\Delta _{1m_{z}}(k)$. The
low energy behavior of background phase shift $\delta _{1}^{\mathrm{(bg)}%
}(k) $ is already known to be \cite{JMP1,gao1}
\begin{eqnarray}  \label{delta1bg}
-k{\cot }\delta _{1}^{\mathrm{(bg)}}(k)&=&\frac{1}{\mathcal{V}^{(\mathrm{bg}%
)}}\frac{1}{k^{2}}+\frac{1}{\mathcal{S}^{(\mathrm{bg})}}\frac{1}{k} +\frac{1%
}{\mathcal{R}^{(\mathrm{bg})}}.  \label{delta1bg}
\end{eqnarray}
Therefore, if we can further obtain the low-energy expansion of term $-k{%
\cot }\Delta _{1m_{z}}(k)$, then the expressions of $-k\cot%
\delta_{1m_{z}}(k) $ and $f_{1m_{z}}(k)$ can be calculated straightforwardly.

\subsection{The low-energy expansion of $-k{\cot }\Delta _{1m_{z}}(k)$}

In this subsection we investigate the expression of $-k{\cot }\Delta
_{1m_z}(k)$ in the limit $k\rightarrow 0$. To this end, we need to expand
both the numerator and the denominator of (\ref{delta1}) in the low-energy
limit.

In this paper we assume the background scattering volume in the open channel
is finite. It can be proved that (Appendix C), in this case the function $%
g_{m_{z}}(k^{2})$ can be expanded as%
\begin{eqnarray}
g_{m_{z}}(k^{2})=g_{m_{z}}^{(0)}+g_{m_{z}}^{(2)}k^{2}+O(k^{3})  \label{gk2}
\end{eqnarray}%
with $g_{m_{z}}^{(2)}\leq0$.

On the other hand, due to the long-range nature of the van der Waals
potential, the partial wave scattering state $|\psi _{k1m_{z}}^{(\mathrm{bg}%
+)}\rangle $ is not an analytical function of the incident momentum $k$ in
the neighborhood of $k=0$ \cite{taylor}. To investigate the low-energy
behavior of $|\psi _{k1m_{z}}^{(\mathrm{bg}+)}\rangle $ and then the
denominator of (\ref{delta1}), we separate the non-analytical part of $|\psi
_{k1m_{z}}^{(\mathrm{bg}+)}\rangle $ by introducing the background Jost
function $\mathscr{J}(k)$ \cite{taylor} defined as%
\begin{eqnarray}
\langle \vec{r}|\psi _{k1m_{z}}^{(\mathrm{bg}+)}\rangle =i^{l}\frac{1}{\hbar %
\mathscr{J}(k)}(\frac{m}{\pi k})^{\frac{1}{2}}\frac{1}{r}Y_{1}^{m_{z}}(\hat{r%
})\tilde{F}_{k1m_{z}}^{(\mathrm{bg})}(r).\label{jost1}
\end{eqnarray}
Here $\tilde{F}_{k1m_{z}}^{(\mathrm{bg})}(r)$ is the canonical solution of
the radial equation%
\begin{eqnarray}
\left(
-\frac{d^{2}}{dr^{2}}+V^{(\mathrm{{{bg})}}}(r)+\frac{2}{r^{2}}\right)
\tilde{F}_{k1m_{z}}^{(\mathrm{bg})}(r)=k^{2}\tilde{F}_{k1m_{z}}^{(\mathrm{bg}%
)}(r)
\end{eqnarray}
with boundary condition%
\begin{eqnarray}
\tilde{F}_{k1m_{z}}^{(\mathrm{bg})}(r\rightarrow 0)\rightarrow \hat{\jmath}%
_{1}(kr)
\end{eqnarray}
where
\begin{eqnarray}
\hat{\jmath}_{1}(x)=\frac{\sin x}{x}-\cos x\label{j1}
\end{eqnarray}
is the first-order regular Riccati-Bessel function \cite{taylor}.
According
to the standard scattering theory \cite{taylor}, $\tilde{F}_{k1m_{z}}^{(%
\mathrm{bg})}(r)$ is an analytical function of $k$, and can be expanded as a
Talyor series of $k$:
\begin{eqnarray}
\tilde{F}_{k1m_{z}}^{(\mathrm{bg})}(r)=\frac{1}{(2n)!}\sum_{n=1}^{\infty
}\left. \frac{d^{2n}}{dk^{2n}}\tilde{F}_{k1m_{z}}^{(\mathrm{bg}%
)}(r)\right\vert _{k=0}k^{2n}.\label{ftuter}
\end{eqnarray}
It is pointed out that, all the odd order terms of the above Taylor
series are exactly zero \cite{taylor}. Thus the non-analytical part
of $|\psi _{k1m_{z}}^{(\mathrm{bg}+)}\rangle $ is included in the
term with the Jost function $\mathscr{J}(k)$.

Substituting Eqs. (\ref{ftuter}), (\ref{jost1}) and (\ref{gk2}) into Eq. (%
\ref{delta1}), we find that in the low-energy limit the factor $-k{\cot }%
\Delta _{1m_{z}}(k)$ takes the form
\begin{eqnarray}
-k{\cot }\Delta _{1m_{z}}(k) &=&\frac{1}{\mathcal{V}^{(\Delta )}(B;k;m_{z})}%
\frac{1}{k^{2}}+\frac{1}{\mathcal{R}^{(\Delta )}(B;k;m_{z})}.  \nonumber
\label{kcot2} \\
&&
\end{eqnarray}%
Here we have the $k$-dependent parameters:%
\begin{eqnarray}
\mathcal{V}^{(\Delta )}(B;k;m_{z}) &=&-|\mathscr{J}(k)|^{-2}\frac{\pi
w_{m_{z}}}{\mu _{\mathrm{res}}}\frac{1}{B-B_{0}}; \\
\mathcal{R}^{(\Delta )}(B;k;m_{z}) &=&|\mathscr{J}(k)|^{-2}\pi
w_{m_{z}}\times   \nonumber \\
&&\left[ \left( \frac{\hbar ^{2}}{m}-g_{m_{z}}^{(2)}\right) -\frac{%
w_{m_{z}}^{\prime }}{w_{m_{z}}}\mu _{\mathrm{res}}\left( B-B_{0}\right) %
\right] ^{-1}.  \nonumber \\
&&  \label{rdet}
\end{eqnarray}%
with the parameters $B_{0}$ and $w_{m_{z}}$ defined by%
\begin{eqnarray}
B_{0} &=&B_{\mathrm{res}}-g_{m_{z}}^{(0)}/\mu _{\mathrm{res}}; \\
w_{m_{z}} &=&\frac{1}{6}\frac{d^{3}}{dk^{3}}\left[ |\langle \phi _{\mathrm{%
res}}^{(m_{z})}|W|\psi _{k1m_{z}}^{(\mathrm{bg}+)}\rangle |^{2}\times |%
\mathscr{J}(k)|^{2}\right] _{k=0}  \nonumber \\
w_{m_{z}}^{\prime } &=&\frac{1}{120}\frac{d^{5}}{dk^{5}}\left[ |\langle \phi
_{\mathrm{res}}^{(m_{z})}|W|\psi _{k1m_{z}}^{(\mathrm{bg}+)}\rangle
|^{2}\times |\mathscr{J}(k)|^{2}\right] _{k=0}.  \nonumber \\
&&
\end{eqnarray}

\subsection{The low-energy p-wave scattering amplitude}

In the above subsection we get the expansion (\ref{kcot2}) of the factor $-k{%
\cot }\Delta _{1m_{z}}(k)$. Substituting Eq. (\ref{kcot2}), (\ref{delta1bg})
and (\ref{delta1a}) into (\ref{fdelta}), we finally get the low-energy
behavior of the p-wave scattering amplitude $f_{1m_{z}}(k)$ in presence of a
PMFR:
\begin{eqnarray}
&&f_{1m_{z}}(k)  \nonumber \\
&=&-\frac{1}{ik+\frac{1}{\mathcal{V}^{\mathrm{eff}}(k;B;m_{z})k^{2}}+\frac{1%
}{\mathcal{S}^{\mathrm{eff}}(k;B;m_{z})k}+\frac{1}{\mathcal{R}^{\mathrm{eff}%
}(k;B;m_{z})}}  \nonumber \\
&&  \label{f1}
\end{eqnarray}%
where the $k$-dependent scattering parameters are given by%
\begin{eqnarray}  \label{v2effa}
\mathcal{V}^{\mathrm{eff}} &=&\mathcal{V}^{\mathrm{(bg)}}\left( 1-\frac{%
b_{m_{z}}}{B-B_{0}}|\mathscr{J}(k)|^{-2}\right) ;  \label{v0eff} \\
\mathcal{S}^{\mathrm{eff}} &=&\frac{\mathcal{S}^{\mathrm{(bg)}}}{\mathcal{V}%
^{\mathrm{{(bg)}2}}}\mathcal{V}^{\mathrm{eff}2};  \label{v1eff} \\
\frac{1}{\mathcal{R}^{\mathrm{eff}}} &=&\frac{1}{\mathcal{R}^{(\Delta )}}%
\left( 1-2x+x^{2}\right) +\frac{1}{\mathcal{R}^{(\mathrm{bg})}}x^{2}+\frac{%
\mathcal{V}^{(\mathrm{bg})}}{\mathcal{S}^{(\mathrm{bg})2}}(x^{2}-x^{3}).
\nonumber \\
\end{eqnarray}%
Here the parameters $b_{m_{z}}$ and $x$ are defined as%
\[
b_{m_{z}}=\frac{\pi w_{m_{z}}}{\mathcal{V}^{\mathrm{(bg)}}\mu _{\mathrm{res}}%
};\ x=\frac{\mathcal{V}^{\mathrm{(bg)}}}{\mathcal{V}^{\mathrm{eff}}}.
\]

So far we have obtained the low-energy expression of the p-wave scattering
amplitude $f_{1m_{z}}(k)$ in the case of PMFR. With the help of the
scattering theory, we obtain the general expressions (\ref{f1}-\ref{v2effa})
for the scattering amplitude $f_{1m_{z}}(k)$ as well as the scattering
parameters $(\mathcal{V}^{\mathrm{eff}},\mathcal{S}^{\mathrm{eff}},\mathcal{R%
}^{\mathrm{eff}})$, which are formulated in terms of the background
scattering parameters $(\mathcal{V}^{(\mathrm{bg})},\mathcal{S}^{(\mathrm{bg}%
)},\mathcal{R}^{(\mathrm{bg})})$ and the magnetic field. It is pointed out
that, although due to the long-range nature of the van der Waals potential
we can not express the denominator of $f_{1m_{z}}(k)$ as an Laurent series
with $k$-independent coefficients, we successfully include all the $k$%
-dependence of the parameters $(\mathcal{V}^{\mathrm{eff}},\mathcal{S}^{%
\mathrm{eff}},\mathcal{R}^{\mathrm{eff}})$ into the Jost function $%
\mathscr{J}(k)$.

Eq. (\ref{v0eff}) clearly show the effect of PMFA that the
scattering volume $V^{\mathrm{eff}}$ diverges under the magnetic
field $B=B_{0}$. In the realistic cold atom gases, for the creation
of an observable effects with $p$-wave interaction, the scattering
volume $|V^{\mathrm{eff}}|$ should be large enough. Particularly,
$|V^{\mathrm{eff}}|^{1/3}$ should be much larger than the van der
Waals length so that in the BCS region, the transition temperature
$T_{c}\sim (E_{F}/k_{B})\exp [-\pi /(2k_{B}^{3}V^{\mathrm{eff}})]$
\cite{bohn} of superfluid is realizable and in the BEC region the
binding energy of the $p$-wave Feshbach molecule be roubst with
respect to the detail of the atom-atom interaction potential.

In the end of this section, we consider the dependence of the
effective range $R^{\mathrm{eff}}$ on the magnetic field $B$.
According to Eq. (\ref{v2effa}), $R^{\mathrm{eff}}$ depends on $B$
through the ratio $x$ between $V^{\mathrm{(bg)}}$ and
$V^{\mathrm{eff}}$, and the quantatity $R^{(\Delta )}(B;k;m_{z})$.
In the cold gases of $^{6}$Li and $^{40}$K, the background
scattering volumes $V^{\mathrm{(bg)}}$ are of the order of
$(10^{5}-10^{6})a_{0}^{3}$. According to our above discussion, they
are too small for the creation of $p$-wave superfluids \cite{bohn}.
Therefore in these systems the strong enough $p$-wave interactions
can only be obtained in the resonance region with
$|V^{\mathrm{eff}}|>>|V^{\mathrm{(bg)}}|$ or $x<<1,$ which implies
$R^{\mathrm{eff}}\approx R^{(\Delta )}(B;k;m_{z})$. On the other
hand, according to Eq. (\ref{rdet}), the dependence of $R^{(\Delta
)}(B;k;m_{z})$ on $B$ is significant when the magnetic field is far
away enough from from the resonant point $B_{0}$ so that the factor
$\left\vert w_{m_{z}}^{\prime }\mu _{\mathrm{res}}\left(
B-B_{0}\right) /w_{m_{z}}\right\vert $ is comparable or larger than
$|\hbar^2/m-g_{m_{z}}^{(2)}|$. The values of $w_{m_{z}}^{\prime }$
and $w_{m_{z}}$ are not available for $^{6}$Li and $^{40}$K.
Nevertheless, the binding energies of the $p$-wave Feshbach
molecules are measured to be linear functions of the magentic field
\cite{expK40Molb,expLi6Mol} in the region with large enough $p$-wave
scattering volumes ($V^{\mathrm{eff}}\gtrsim 10^{7}a_{0}^{3}$). This
observation shows that in these regions the term $\left\vert
w_{m_{z}}^{\prime }\mu _{\mathrm{res}}\left( B-B_{0}\right)
/w_{m_{z}}\right\vert $ is negligible and the effective range
$R^{\mathrm{eff}}$ can be approximated as a $B$-independent constant
$R^{(\Delta )}(0;k;m_{z})$.

\section{The applicability of the effective-range theory}

In the above section, we have obtained the expression (\ref{f1}) of the
p-wave scattering amplitude $f_{1m_{z}}(k)$ in the region near the point of
PMFR. It is apparent that, this expression is different from the one (\ref%
{1a}) given by the effective-range theory in the following two senses:

1. In the standard effective-range theory, the scattering volume $\mathcal{V}
$ and effective-range $\mathcal{R}$ are independent on the incident momentum
$k$. Nevertheless, in the expression (\ref{f1}) the scattering parameters $(%
\mathcal{V}^{\mathrm{eff}},\mathcal{S}^{\mathrm{eff}},\mathcal{R}^{\mathrm{%
eff}})$ depend on $k$ through the Jost function $\mathscr{J}(k)$.

2. The term $1/[\mathcal{S}^{\mathrm{eff}}(k;B;m_{z})k]$ cannot be included
in the effective-range theory.

It is apparent that, if under some condition the scattering amplitude (\ref%
{f1}) can be approximated as%
\begin{equation}
f_{1m_{z}}(k)\approx -\frac{1}{ik+\frac{1}{\mathcal{V}^{\mathrm{eff}%
}(0;B;m_{z})k^{2}}+\frac{1}{\mathcal{R}^{\mathrm{eff}}(0;B;m_{z})}},
\label{f1er}
\end{equation}%
i.e., both the $k$-dependence of $|\mathscr{J}(k)|^{2}$ and the term $1/[%
\mathcal{S}^{\mathrm{eff}}(k;B;m_{z})k]$ can be neglected, the behavior of
the system would be approximately described by the effective-range theory.
In this section, we investigate the conditions for the approximation (\ref%
{f1er}) or (\ref{1g}), or the applicability of effective-range theory. We
will consider the importance of the term $1/[\mathcal{S}^{\mathrm{eff}%
}(k;B;m_{z})k]$ and the $k$-dependence of the Jost function respectively.

\subsection{The $k$-dependence of the scattering parameters}

In this subsection we search the sufficient condition for the ignorance of
the $k$-dependence of the scattering parameters $(\mathcal{V}^{\mathrm{eff}},%
\mathcal{S}^{\mathrm{eff}},\mathcal{R}^{\mathrm{eff}})$. As we have
discussed above, the $k$-dependence of the scattering parameters comes from
the mode square of the Jost function $\mathscr{J}(k)$. In the ultracold
gases of the fermionic atoms, the maximum value of the relative momentum of
two atoms is on the order of the Fermi momentum $k_F$. Therefore, the
importance of the $k$-dependence of the parameters parameters $(\mathcal{V}^{%
\mathrm{eff}},\mathcal{S}^{\mathrm{eff}},\mathcal{R}^{\mathrm{eff}})$ can be
described by the factor
\begin{eqnarray}
r_{1}=\frac{|\mathscr{J}(k_{F})|^{-2}-|\mathscr{J}(0)|^{-2}}{|\mathscr{J}%
(0)|^{-2}}.  \label{r2}
\end{eqnarray}
Obviously, when $r_{1}<<1$, we can replace $|\mathscr{J}(k)|^{-2}$ with $|%
\mathscr{J}(0)|^{-2}$ and neglect the $k-$ dependence of the parameters $(%
\mathcal{V}^{\mathrm{eff}},\mathcal{S}^{\mathrm{eff}},\mathcal{R}^{\mathrm{%
eff}})$.

\begin{figure}[tbp]
\includegraphics[bb=30bp 232bp 517bp 574bp,clip,width=7.5cm]{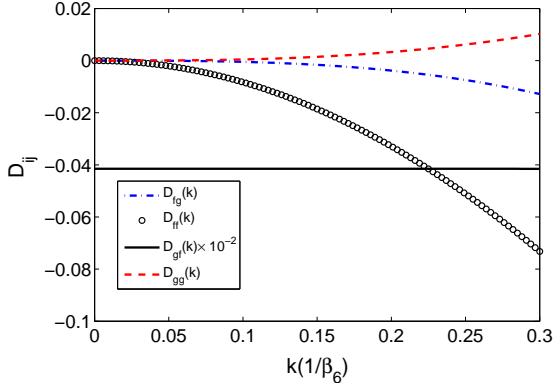}
\caption{(color online) The functions $D_{fg}(k)$ (blue dash-dotted line), $%
D_{ff}(k)$ (open circle), $D_{gf}(k)$ (black solid line) and $D_{gg}(k)$
(red dashed line) in Eq. (\protect\ref{jost}).}
\end{figure}

To investigate the behavior of the the ratio $r_1$, we first calculate the
Jost function $\mathscr{J}(k)$. By means of the quantum defect theory \cite%
{gao3}, we can obtain the expression of $|\mathscr{J}(k)|^{-2}$ (see
Appendix D):
\begin{eqnarray}  \label{jost}
&&|\mathscr{J}(k)|^{-2}=\alpha ^{-2}\beta _{6}^{3}\frac{\pi}{2}\times
\nonumber \\
&&\left[ \left( D_{ff}(k)-K_{l=1}^{0}D_{gf}(k)\right) ^{2}+\left(
D_{fg}(k)-K_{l=1}^{0}D_{gg}(k)\right) ^{2}\right]^{-1}  \nonumber \\
\end{eqnarray}
where $\alpha $ is a $k$-independent coefficient and
$D_{ij}(k)=(k\beta _{6})^{3/2}Z_{ij}(k)$ $(i,j=f,g)$ with
$Z_{ij}(k)$ defined in \cite{gao2}. In Fig. 2 we plot the functions
$D_{ij}(k)$ in the low-energy case.

The parameter $K_{l=1}^0$ is denoted as $K_{l=1}^0$ is related to the
background scattering parameters \cite{gao1}. Expanding the p-wave phase
shift in Eq.(7) of Ref. \cite{gao1}, we can express $(\mathcal{V}^{(\mathrm{%
bg)}},\mathcal{S}^{(\mathrm{bg)}})$ in terms of $K_{l=1}^0$:
\begin{eqnarray}
\mathcal{V}^{(\mathrm{bg)}}&=&-\frac{(1+K_{l=1}^0)\pi
}{18K_{l=1}^0\Gamma
\lbrack 3/4]^{2}}\beta _{6}^{3};  \label{v0bg} \\
\mathcal{S}^{(\mathrm{bg)}}&=&-\frac{35(1+K_{l=1}^0)^{2}\pi }{%
324(K_{l=1}^{0})^2\Gamma \lbrack 3/4]^{4}}\beta _{6}^{2}.  \label{v1bg}
\end{eqnarray}

The above expression shows that, when $K_{l=1}^{0}\gtrsim
\pi/(18\Gamma[3/4]^2)\sim0.1$, we have
$\mathcal{V}^{(\mathrm{bg)}}\sim \beta_6^3$ and the background
scattering potential is far away from the shape resonance; when
$K_{l=1}^{0}$ is much smaller than $0.1$, the background potential
is in the shape resonance region which gives
$\mathcal{V}^{(\mathrm{bg)}}>> \beta_6^3$.

Now we consider the features of the ratio $r_1$, which
 is determined by the parameter $K_{l=1}^{0}$. Fig. 2 shows that in the low energy case with $k\beta_6<< 1$, the
function $D_{gf}(k)$ is almost a $k$-independent constant and much
larger than the other three D-functions. Therefore, if the parameter
$K_{l=1}^{0}$ is large or the background scattering potential in the
open channel is far away from the shape resonance, then according to
Eq. (\ref{jost}), the Jost function $|\mathscr{J}(k)|$ is dominated
by the term with $K_{l=1}^{0}D_{gf}(k)$. In this case the variation
of $|\mathscr{J}(k)|^{-2}$ with respect to $k$ is negligible and we
have $r_{1}<<1$. On the other hand, if $K_{l=1}^{0}$ is close to
zero and the background scattering potential is close to the shape
resonance, then $|\mathscr{J}(k)|$ becomes a rapid changing function
of $k$ and the ratio $r_1$ would be significant.

The above argument is quantitatively verified by Fig. 2, where the
ratio $r_{1}$ is plotted as functions of $k$ with respect different
values of $K_{l=1}^0$ or $\mathcal{V}^{(\mathrm{bg)}}$. It is
clearly shown that, if the fermonic momentum $k_{F}\lesssim 0.1\beta
_{6}^{-1}$, then the ratio $r_{1}$ and the $k$-dependence of the
scattering parameters can be neglected when the background potential
is far enough from the shape resonance so that
$\mathcal{V}^{(\mathrm{bg)}}\lesssim \beta _{6}^{3}$. If
$k_{F}\lesssim 0.01\beta _{6}^{-1}$, this restriction can be further
relaxed to $\mathcal{V}^{(\mathrm{bg)}}\lesssim 10\beta _{6}^{3}$.

\begin{figure}[tbp]
\includegraphics[bb=44bp 218bp 517bp 573bp,clip,width=7.5cm]{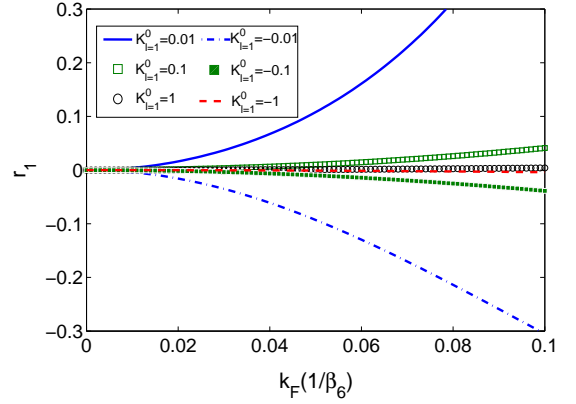}
\caption{(color online) Ratio $r_{1}$ defined in Eq. (\protect\ref{r2}) as a
function of the Fermi momentum $k_{F}$ with $K_{l=1}^{0}=0.01$ ($\mathcal{V}%
^{(\mathrm{bg)}}=-11.79\protect\beta _{6}^{3}$, blue solid line), $%
K_{l=1}^{0}=0.1$ ($\mathcal{V}^{(\mathrm{bg)}}=-1.27\protect\beta
_{6}^{3}$,
green empty square), $K_{l=1}^{0}=1$ ($\mathcal{V}^{(\mathrm{bg)}}=-0.23%
\protect\beta _{6}^{3}$, black empty circle), $K_{l=1}^{0}=-0.01$ ($\mathcal{%
V}^{(\mathrm{bg)}}=11.51\protect\beta _{6}^{3}$, blue dash-dotted line), $%
K_{l=1}^{0}=-0.1$ ($\mathcal{V}^{(\mathrm{bg)}}=10.4\protect\beta _{6}^{3}$%
, green filled square), $K_{l=1}^{0}=-1$ ($\mathcal{V}^{(\mathrm{bg)}}=0$,
red dashed line).}
\label{fig1}
\end{figure}

\subsection{The importance of the term $1/[\mathcal{S}^{\mathrm{eff}%
}(k;B;m_{z})k]$}

Now we discuss the importance of the term $1/[\mathcal{S}^{\mathrm{eff}%
}(k;B;m_{z})k]$. Since the purpose of this paper is to obtain the sufficient
condition for the effective-range theory, or the approximation (\ref{f1er}),
for simplicity, in this subsection we assume the condition $r_{1}<<1$
obtained in the above section is already met, and the $k$-dependence of the
coefficients $(\mathcal{V}^{\mathrm{eff}},\mathcal{S}^{\mathrm{eff}},%
\mathcal{R}^{\mathrm{eff}})$ can be neglected.

In the BEC side of the PMFR where $B<B_{0}$, since $g_{m_{z}}^{(2)}<0$, the
parameters $\mathcal{R}^{\mathrm{eff}}$ and $\mathcal{V}^{\mathrm{eff}}$
have the same sign. In that case, if the absolute value of $1/(\mathcal{S}^{%
\mathrm{eff}}k)$ is much smaller than the one of $1/(\mathcal{V}^{\mathrm{eff%
}}k^{2})$, then it would be also much smaller than $1/(\mathcal{V}^{\mathrm{%
eff}}k^{2})+1/\mathcal{R}^{\mathrm{eff}}$, and can be neglected. It is
obvious that, in the limit $k\rightarrow 0$, the term $1/(\mathcal{S}^{%
\mathrm{eff}}k)$ would be much smaller than $1/(\mathcal{V}^{\mathrm{eff}%
}k^{2})$. Therefore the importance of the term $1/(\mathcal{S}^{\mathrm{eff}%
}k)$ is actually determined by the ratio $r_{2}$ between the two terms for
the upper limit of the relative momentum $k_{F}$:
\begin{eqnarray}
r_{2}=\left\vert \frac{1/[\mathcal{S}^{\mathrm{eff}}(0;B;m_{z})k_{F}]}{1/[%
\mathcal{V}^{\mathrm{eff}}(0;B;m_{z})k_{F}^{2}]}\right\vert .
\end{eqnarray}
When $r_{2}$ is much smaller than unity, we can neglect the term $1/(%
\mathcal{S}^{\mathrm{eff}}k)$ . If $r_{2}$ is comparable or larger than
unity, the term $1/(\mathcal{S}^{\mathrm{eff}}k)$ would be necessary for the
theory.

The straightforward calculation with Eqs. (\ref{v0eff}), (\ref{v1eff}), (\ref%
{v0bg}) and (\ref{v1bg}) yields
\begin{equation}
r_{2}=\frac{\pi }{35}\frac{\beta _{6}^{3}}{|\mathcal{V}^{\mathrm{eff}%
}(k_{F};B;m_{z})|}(\beta _{6}k_{F}).  \label{r1new}
\end{equation}%
In the practical cold atom systems we have $\beta _{6}k_{F}<<1$. Therefore,
in the resonance region with $\mathcal{V}^{\mathrm{eff}}\gtrsim \beta
_{6}^{3}$, the ratio $r_{2}$ in Eq. (\ref{r1new}) is much smaller than
unity, and then the term $1/(\mathcal{S}^{\mathrm{eff}}k),$ is negligible.

In the BEC side of the resonance with $B>B_{0}$, the parameters $\mathcal{V}%
^{\mathrm{eff}}$ and $\mathcal{R}^{\mathrm{eff}}$ have different signs. In
that case, there is a speical momentum%
\begin{equation}
k_{\ast }=\sqrt{-\frac{\mathcal{R}^{\mathrm{eff}}(0;B;m_{z})}{\mathcal{V}^{%
\mathrm{eff}}(0;B;m_{z})}}  \label{kstar}
\end{equation}%
which makes the terms\ $1/[\mathcal{V}^{\mathrm{eff}}k_{\ast }^{2}]$ and $1/[%
\mathcal{R}^{\mathrm{eff}}]$ cancel with each other or%
\begin{equation}
\frac{1}{\mathcal{V}^{\mathrm{eff}}(0;B;m_{z})k_{\ast }^{2}}+\frac{1}{%
\mathcal{R}^{\mathrm{eff}}(0;B;m_{z})}=0.
\end{equation}

Therefore, if the atomic relative momentum $k$ is far away from $k_{*}$, the
absolute value of $1/(\mathcal{R}^{\mathrm{eff}}k^{2})$ would be quite
different with the one of $1/\mathcal{V}^{\mathrm{eff}}$. In that case we
can still neglect the term $1/(\mathcal{S}^{\mathrm{eff}}k)$ under the
condition $r_2<<1$ or $|1/(\mathcal{S}^{\mathrm{eff}}k)|<<|1/(\mathcal{R}^{%
\mathrm{eff}}k^2)|$.

If the atomic relative momentum $k$ is in the neighborhood of $k_{\ast }$
and the terms $1/(\mathcal{R}^{\mathrm{eff}}k_{\ast }^{2})$ is canceled with
$1/\mathcal{V}^{\mathrm{eff}}$, the scattering amplitude (\ref{f1}) can be
expressed as%
\begin{equation}
f_{1m_{z}}(k)=-\frac{1}{ik_{\ast }+\frac{1}{\mathcal{S}^{\mathrm{eff}%
}(0;B;m_{z})k_{\ast }}}.
\end{equation}%
In that case, if the absolute value of $1/\left[ \mathcal{S}^{\mathrm{eff}%
}\left( 0;B;m_{z}\right) k_{\ast }\right] $ is much smaller than $k_{\ast }$%
, we an also neglect the term with $\mathcal{S}^{\mathrm{eff}}$, even in the
neighborhood of $k_{\ast }$. We define a parameter $r_{3}$ as
\[
r_{3}=\frac{1}{\left\vert \mathcal{S}^{\mathrm{eff}}\left( 0;B;m_{z}\right)
k_{\ast }^{2}\right\vert }.
\]%
Then the term with $\mathcal{S}^{\mathrm{eff}}$ can be neglected when $%
r_{2,3}<<1$. A further calculation with Eqs. (\ref{kstar}), (\ref{v0eff}), (%
\ref{v1eff}), (\ref{v0bg}) and (\ref{v1bg}) implies that
\begin{equation}
r_{3}=\frac{\pi \beta _{6}^{4}}{35|\mathcal{V}^{\mathrm{eff}}\left(
0;B;m_{z}\right) \mathcal{R}^{\mathrm{eff}}\left( 0;B;m_{z}\right) |}.
\label{r3}
\end{equation}

As shown above, the condition $r_{3}<<1$ is obtained for the
momentum region $k\sim k_{\ast }$. Since the realistic momentum of
the atomic relative motion takes the value between zero and $k_{F}$,
in the cases with $k_{F}<k_{\ast }$, we can disregard the
restriction of the ratio $r_{3}$, and use effective-range theory
under the condition $r_{1,2}<<1$ in both the BEC and the BCS sides
of the resonance.

\subsection{Summary}

In summary, the general sufficient conditions for the
effective-range theory in the BCS side of the resonance can be
summarized as
\begin{equation}
r_{1},r_{2},r_{3}<<1,  \label{sc1}
\end{equation}while the ones for the BEC side are
\begin{equation}
r_{1},r_{2}<<1.  \label{sc2}
\end{equation}From the definition of the ratios $r_{1},r_{2}$ and $r_{3}$, we notice that
in the realistatic cold gases of Fermi atoms, the crucial factors
for the usage of effective range theory is the background $p$-wave
scattering volume $\mathcal{V}^{\mathrm{(bg)}}$ and the
$B$-dependence of the factor $\mathcal{R}^{(\Delta )}\left(
0;B;m_{z}\right) $. If the the background $p$-wave scattering is far
away from the shape resonance so that
$\mathcal{V}^{\mathrm{(bg)}}\sim \beta _{6}^{3}\sim (100a_{0})^{3}$
then according to our previous discussions and Eqs. (\ref{r1new}),
the conditions $r_{1},r_{2}<<1$ can be satisfied in the region
$\mathcal{V}^{\mathrm{eff}}\gtrsim 10\beta _{6}^{3}\sim
10^{7}a_{0}^{3}$ where the $p$-wave interaction is strong enough for
the creation of $p$-wave superfluids. In that region we also have
$\mathcal{R}^{\mathrm{eff}}\left( 0;B;m_{z}\right)
=\mathcal{R}^{(\Delta )}\left( 0;B;m_{z}\right) $. If
$\mathcal{R}^{(\Delta )}\left( 0;B;m_{z}\right) $ can be further
approximated as a $B$-independent constant which is of the order
$\beta _{6}$, then the condition $r_{3}<<1$ can also be satisfied,
and the effective range theory can be used a good approximation for
the realistic scattering amplitude. In the following subsection we
show that the PMFRs in the cold gases of $^{40}$K and $^{6}$Li are
just of this case.

\subsection{Discussion for the cold gases of $^{40}$K and $^{6}$Li}

In the above subsections we obtained the sufficient conditions (\ref{sc1}), (%
\ref{sc2}) of the effective-range theory for the p-wave scattering
amplitudes of polarized fermonic atoms near a PMFR. In this subsection, with
the help of the conditions, we perform a discussion on the usage of
effective-range theory in the ultracold gases of $^{40}$K and $^{6}$Li.

For the ultracold gas with $^{40}$K atoms in the state $|9/2,-7/2\rangle $,
we have $C_{6}=3897$(a.u.) \cite{c6k} and $\mathcal{V}^{\mathrm{(bg)}%
}=-10^{6}a_{0}^{3}$ \cite{expK40Dipo}. These parameters leads $\beta
_{6}=130a_{0}$ and $K_{l=1}^{0}=-0.16$. If the Fermi temperature $T_{F}=1\mu
$K, then we have $k_{F}\beta _{6}=0.06$. The straightforward calculation
shows that $r_{1}=0.01$. Therefore $k$-dependence of the scattering
parameters can be safely neglected. The p-wave Feshbach resonance for the
states with $m_{l}=\pm 1$ occurs at $B_{0}=198.37$G with width $\Delta B=25$%
G and effective-range $\mathcal{R}^{\mathrm{eff}}=47.2a_{0}$. The resonance
for the states with $m_{l}=0$ occurs at $B_{0}=198.85$G with width $\Delta
B=22$G and effective-range $\mathcal{R}^{\mathrm{eff}}=46.2a_{0}$ \cite%
{expK40Dipo}. According to these data we have $r_{3}<0.02$ when $k_{\ast
}<k_{F}$. Then effect from the ratio $r_{3}$ is also negligible. Therefore
the sufficient condition for the usage of effective-range simply becomes $%
r_{2}<<1$. Further calculation shows that $r_{2}\leq 0.013$ when $|\mathcal{V%
}^{\mathrm{eff}}|\geq |\mathcal{V}^{\mathrm{(bg)}}|$. Then the
effective-range approximation (\ref{f1er}) is applicable for $^{40}$K atoms
in the state $|9/2,-7/2\rangle $ in the whole region of PMFR with $|\mathcal{%
V}^{\mathrm{eff}}|\geq |\mathcal{V}^{\mathrm{(bg)}}|$. The condition for the
effective-range approximation is broken only in the small region $220.5%
\mathrm{G}<B<221$G ($m_{l}=0$) or $223\mathrm{G}<B<223.7$G ($m_{l}=\pm 1$)
where we have $|\mathcal{V}^{\mathrm{eff}}|\leq 0.005\beta _{6}^{3}$ or $%
r_{2}\geq 1$.

Now we consider the gas with $^{6}$Li atoms in the ground hyperfine state $%
|F=1;m_{F}=1\rangle $. In that case we have $C_{6}=1393$(a.u.) \cite%
{li6parameter,XXXGao} and $\mathcal{V}^{\mathrm{(bg)}}=-(35.3a_{0})^{3}$.
These parameters leads $\beta _{6}=62a_{0}$ and $K_{l=1}^{0}=-0.38$. If the
Fermi temperature $T_{F}=1\mu $K, we have $k_{F}\beta _{6}=0.01$ which
implies $r_{1}=2\times 10^{-4}$. Then similar as above, the effective range
approximation (\ref{f1er}) is also applicable for $^{6}$Li atoms in the
whole region of PMFR with $|\mathcal{V}^{\mathrm{eff}}|\geq |\mathcal{V}^{%
\mathrm{(bg)}}|$.

\section{Conclusion}

In this work we obtain the explicit expression of the p-wave scattering
amplitude of two ultracold spin polarized fermonic atoms near the p-wave
Feshbach resonance. We show that due to the long rang nature of the van der
Waals potential, the scattering amplitude is explicitly described by Eq. (%
\ref{1c}) in the low-energy case. With the help of the quantum
defect
theory, we formulate all the scattering parameters $(\mathcal{V}^{\mathrm{eff%
}},\mathcal{S}^{\mathrm{eff}},\mathcal{R}^{\mathrm{eff}})$ in terms of the
background parameters and the inter-channel coupling.

Based on this result, we discussed the applicability of the
effective-range theory, or the Eq. (\ref{1d}) as an approximation of
the exact scattering amplitude. We show that, in the BEC side of the
resonance, the sufficient
conditions of the effective-range theory can be quantitatively described as $%
r_{1},r_{2}<<1$ while in the BCS side the conditions become $%
r_{1},r_{2},r_{3}<<1$, where $r_{1},r_{2},r_{3}$ are defined in Eqs. (\ref%
{r1new}), (\ref{r2}), and (\ref{r3}). The applicability of the
effective-range theory for the ultracold gases of $^{40}$K and $^{6}$Li are
examined with our results. The effective-range theory is shown to be a good
approximation in both of the two cases in the resonance regime where the
absolute value of the scattering volume is equal to or larger then the
background one.

\appendix%\appendixpage
\addcontentsline{toc}{section}{Appendices}\markboth{APPENDICES}{}
\begin{subappendices}

\section{The p-wave phase shift with PMFR}

In this appendix we post the derivation of the p-wave phase shift $\delta
_{1m_{z}}(k)$ in Eqs. (\ref{delta1a}) and (\ref{delta1}). Our calculation is
based on the method in Ref. \cite{julian}. We begin from the scattering
state $|\Phi _{\vec{k}}^{(+)}\rangle $ in Eq. (\ref{phik}). According to the
scattering theory \cite{taylor}, the open channel component $|\phi _{\vec{k}%
}^{(\mathrm{op}+)}\rangle$ and closed channel component $|\phi _{\vec{k}}^{(%
\mathrm{cl})}\rangle$ of $|\Phi _{\vec{k}}^{(+)}\rangle $ satisfy the two
channel Lippmman-Schwinger equation%
\begin{widetext}
\begin{eqnarray}
|\phi _{\vec{k}}^{(\mathrm{op}+)}\rangle &=&|\vec{k}\rangle
+G_{0}^{(+)}(k^{2})V^{(\mathrm{bg})}|\phi
_{\vec{k}}^{(\mathrm{op}+)}\rangle+G_{0}^{(+)}(k^{2})W|\phi _{\vec{k}}^{(\mathrm{cl}+)}\rangle;    \label{lp1a} \\
|\phi _{\vec{k}}^{(\mathrm{cl}+)}\rangle &=&G_{0}^{(\mathrm{cl})}(k^{2})V^{(%
\mathrm{cl})}|\phi _{\vec{k}}^{(\mathrm{cl}+)}\rangle+G_{0}^{(\mathrm{cl}%
)}(k^{2})W|\phi _{\vec{k}}^{(\mathrm{op}+)}\rangle  \label{lp1b}
\end{eqnarray}%
\end{widetext}
with the free Green's functions%
\begin{eqnarray}
G_{0}^{(+)}(k^{2}) &=&\frac{1}{\hbar^2k^{2}/m+i0^{+}-\hat{T}}; \\
G_{0}^{(\mathrm{cl})}(k^{2})
&=&\frac{1}{\hbar^2k^{2}/m-\hat{T}-\varepsilon (B)}.
\end{eqnarray}%
We further define the background Green's function $G_{\mathrm{bg}%
}^{(+)}(k^{2})$ and the closed channel Green's function $G_{\mathrm{cl}}$ as%
\begin{eqnarray}
G_{\mathrm{bg}}^{(+)}(k^{2}) &=&\frac{1}{\hbar^2k^{2}/m+i0^{+}-\hat{T}-V^{(\mathrm{bg}%
)}};  \label{gbg} \\
G_{\mathrm{cl}}(k^{2}) &=&\frac{1}{\hbar^2k^{2}/m-\hat{T}-V^{(\mathrm{cl}%
)}-\varepsilon (B)}.  \label{gcl}
\end{eqnarray}%
Then we have the relationships
\begin{eqnarray}
G_{0}^{(+)}(k^{2}) &=&G_{\mathrm{bg}}^{(+)}(k^{2})-G_{\mathrm{bg}%
}^{(+)}(k^{2})V^{(\mathrm{bg})}G_{0}^{(+)}(k^{2});  \label{g01} \\
G_{0}^{(\mathrm{cl})}(k^{2}) &=&G_{\mathrm{cl}}(k^{2})-G_{\mathrm{cl}%
}(k^{2})V^{(\mathrm{cl})}G_{0}^{(\mathrm{cl})}(k^{2}).  \label{g02}
\end{eqnarray}%
Substituting Eqs. (\ref{g01}) and (\ref{g02}) into the last terms of the
right hand side (r.h.s.) of Eqs. (\ref{lp1a}) and (\ref{lp1b}), and using
the Lippmman Schwinger equation (\ref{psibg}) for the background scattering
state $|\phi _{\vec{k}}^{(\mathrm{bg}+)}\rangle $, we get the equation which
relates $|\Phi _{\vec{k}}^{(+)}\rangle $ with the background scattering
state  $|\phi _{\vec{k}}^{(\mathrm{bg}+)}\rangle $ \cite{julian}:
\begin{eqnarray}
|\phi _{\vec{k}}^{(\mathrm{op}+)}\rangle &=&|\phi _{\vec{k}}^{(\mathrm{bg}%
+)}\rangle +G_{\mathrm{bg}}^{(+)}(k^{2})W|\phi _{\vec{k}}^{(\mathrm{cl}%
+)}\rangle  \label{m} \\
|\phi _{\vec{k}}^{(\mathrm{cl}+)}\rangle &=&G_{\mathrm{cl}}(k^{2})W|\phi _{%
\vec{k}}^{(\mathrm{op}+)}\rangle .  \label{n}
\end{eqnarray}

To calculate the p-wave phase shifts $\delta _{1m_{z}}(k)$, we operate the
projection operator $\mathcal{P}_{m_{z}}$ for the manifold $\left(
l=1,L_{z}=m_{z}\right) \ $on both of the two sides of Eqs. (\ref{m}) and (%
\ref{n}). Then we have%
\begin{eqnarray}
\mathcal{P}_{m_{z}}|\phi _{\vec{k}}^{(\mathrm{op}+)}\rangle &=&\mathcal{P}%
_{m_{z}}|\phi _{\vec{k}}^{(\mathrm{bg}+)}\rangle +\mathcal{P}_{m_{z}}G_{%
\mathrm{bg}}^{(+)}(k^{2})W\mathcal{P}_{m_{z}}|\phi _{\vec{k}}^{(\mathrm{cl}%
+)}\rangle  \nonumber \\
&&  \label{mm} \\
\mathcal{P}_{m_{z}}|\phi _{\vec{k}}^{(\mathrm{cl}+)}\rangle &=&|\phi _{%
\mathrm{res}}^{(m_{z})}\rangle \frac{\langle \phi _{\mathrm{res}}^{(m_{z})}|W%
\mathcal{P}_{m_{z}}|\phi _{\vec{k}}^{(\mathrm{op}+)}\rangle }{\hbar^2k^{2}/m-\mu _{%
\mathrm{res}}(B-B_{\mathrm{res}}^{(m_{z})})}.  \label{nn}
\end{eqnarray}%
Here we have used%
\begin{eqnarray}
\mathcal{P}_{m_{z}}G_{\mathrm{bg}}^{(+)}(k^2) &=&\mathcal{P}_{m_{z}}G_{\mathrm{%
bg}}^{(+)}(k^2)\mathcal{P}_{m_{z}}; \\
\mathcal{P}_{m_{z}}G_{\mathrm{cl}}(k^{2}) &=&\mathcal{P}_{m_{z}}G_{\mathrm{cl%
}}(k^{2})\mathcal{P}_{m_{z}}
\end{eqnarray}%
which are guaranteed by the rotational symmetry along the $z$-axis of the
system. We also made the approximation
\begin{eqnarray}
G_{\mathrm{cl}}(k^{2})\approx \sum_{m_{z}}\frac{|\phi _{\mathrm{res}%
}^{(m_{z})}\rangle \langle \phi _{\mathrm{res}}^{(m_{z})}|}{\hbar^2k^{2}/m-\mu _{\mathrm{res}}(B-B_{\mathrm{res}%
}^{(m_{z})})}.
\end{eqnarray}
That is, we only take into account the contribution from the
near-resonance bound state $|\phi _{\mathrm{res}}^{(m_{z})}\rangle$
in the closed channel.

Substituting Eqs. (\ref{nn}) into Eq. (\ref{mm}), we get%
\begin{eqnarray}
\mathcal{P}_{m_{z}}|\phi _{\vec{k}}^{(\mathrm{op}+)}\rangle&=&\mathcal{P}%
_{m_{z}}|\phi _{\vec{k}}^{(\mathrm{bg}+)}\rangle \nonumber\\
&&+G_{\mathrm{bg}%
}^{(+)}(k^2)W|\phi _{\mathrm{res}}^{(m_{z})}\rangle A^{(m_{z})}(B,k^2)\nonumber\\
\label{o}
\end{eqnarray}%
with
\begin{eqnarray}
A^{(m_{z})}(B,k^2)=\frac{\langle \phi _{\mathrm{res}}^{(m_{z})}|W\mathcal{P}%
_{m_{z}}|\phi _{\vec{k}}^{(\mathrm{op}+)}\rangle }{\hbar^2k^{2}/m-\mu _{\mathrm{res}%
}(B-B_{\mathrm{res}}^{(m_{z})})}.  \label{p}
\end{eqnarray}%
Replacing the $\mathcal{P}_{m_{z}}|\phi _{\vec{k}}^{(\mathrm{op}+)}\rangle $
in the r.h.s. of Eq. (\ref{p}) with the r.h.s. of (\ref{o}), we get%
\begin{eqnarray}
&&A^{(m_{z})}(B,k^2)\nonumber\\
&=&\frac{\langle \phi _{\mathrm{res}}^{(m_{z})}|W|\phi _{\vec{k}}^{(\mathrm{%
bg}+)}\rangle }{\hbar^2k^{2}/m-\mu
_{\mathrm{res}}(B-B_{\mathrm{res}}^{(m_{z})})-\langle
\phi _{\mathrm{res}}^{(m_{z})}|WG_{\mathrm{bg}}^{(+)}(E)W|\phi _{\mathrm{res}%
}^{(m_{z})}\rangle }. \nonumber\\
\label{q}
\end{eqnarray}%

Substituting Eq. (\ref{q}) into Eq. (\ref{o}), and
using the asymptotic
expression (\ref{asypsi}) of the scattering state and the definition (\ref%
{parf}) of the partial wave scattering amplitude, we can obtain the p-wave
scattering amplitude%
\begin{widetext}
\begin{eqnarray}
f_{1m_{z}}(k)=f_{1m_{z}}^{\mathrm{(bg)}}(k)-\frac{\pi }{k}\frac{\langle \psi _{k1m_{z}}^{(\mathrm{bg}-)}|W|\phi _{%
\mathrm{res}}^{(m_{z})}\rangle \langle \phi _{\mathrm{res}}^{(m_{z})}|W|\psi
_{k1m_{z}}^{(\mathrm{bg}+)}\rangle }{\hbar^2k^{2}/m-\mu _{\mathrm{res}}(B-B_{\mathrm{res}%
}^{(m_{z})})-\langle \phi _{\mathrm{res}}^{(m_{z})}|WG_{\mathrm{bg}%
}^{(+)}(k^{2})W|\phi _{\mathrm{res}}^{(m_{z})}\rangle }.
\label{fcomplex}
\end{eqnarray}%
\end{widetext}
Here $|\psi _{k1m_{z}}^{(\mathrm{bg}+)}\rangle $ is defined in Eq. (\ref%
{psik1mz}). $|\psi _{k1m_{z}}^{(\mathrm{bg}-)}\rangle $ is defined as
\begin{eqnarray}
|\phi _{\vec{k}}^{(\mathrm{bg}-)}\rangle =\left( \frac{2}{m\hbar k}\right) ^{\frac{1%
}{2}}\sum_{l,m_{z}}|\psi _{k1m_{z}}^{(\mathrm{bg}-)}\rangle Y_{l}^{m_{z}}(%
\hat{k})^{\ast }.  \label{psik1mzn}
\end{eqnarray}%
with%
\begin{eqnarray}
|\phi _{\vec{k}}^{(\mathrm{bg}-)}\rangle =|\vec{k}\rangle +\frac{1}{%
\hbar^2k^{2}/m+i0^{-}-\hat{T}-V^{(\mathrm{bg})}}V^{(\mathrm{{{bg})}}}|\vec{k}\rangle\nonumber\\
\end{eqnarray}%
the background state with in-going boundary condition. In the above
calculation we also used the asymptotic behavior of the background
Green's function:%
\begin{eqnarray}
\lim_{r\rightarrow \infty }\langle \vec{r}|G_{\mathrm{bg}}^{(+)}(k^{2})|\vec{%
r}^{\prime }\rangle =-m\sqrt{\frac{\pi }{2\hbar}}\frac{e^{ikr}}{r}\langle \phi _{k%
\hat{r}}^{(\mathrm{bg}-)}|\vec{r}\rangle
\end{eqnarray}%
with $\hat{r}=\vec{r}/r$.

With straightforward calculation, we can further rewrite the scattering
amplitude $f_{1m_{z}}(k)$ in Eq. (\ref{fcomplex}) as
\begin{eqnarray}
f_{1m_{z}}(k)=f_{1}^{\mathrm{(bg)}}(k)-e^{2i\delta _{1}^{\mathrm{%
(bg)}}(k)}\frac{1}{ik+\mathcal{C}(k)}\label{finalf}
\end{eqnarray}%
with%
\begin{eqnarray}
&&\mathcal{C}(k)=\frac{k}{\pi }\times \frac{1}{|\langle \phi _{\mathrm{res}%
}^{(m_{z})}|W|\psi _{k10}^{(+)}\rangle |^{2}}\times \nonumber \\
&&\left( \hbar^2k^{2}/m-\mu
_{\mathrm{res}}(B-B_{\mathrm{res}}^{(m_{z})})-g_{m_z}(k^2) \right) .
\end{eqnarray}%
In the derivation of Eq. (\ref{finalf}) we have used the
relationship
(Appendix B)%
\begin{eqnarray}
|\psi _{k1m_{z}}^{(-)}\rangle =e^{-2i\delta _{1}^{\mathrm{(bg)}%
}(k)}|\psi _{k1m_{z}}^{(+)}\rangle\label{nppsi}
\end{eqnarray}%
and
\begin{eqnarray}
G_{\mathrm{bg}}^{(+)}(k^{2}) &=&(m\hbar)\int d\vec{k'}\frac{|\phi _{\vec{k'}}^{(%
\mathrm{bg}+)}\rangle \langle \phi _{\vec{k'}}^{(\mathrm{bg}+)}|}{%
k^{2}+i0^{+}-k'^{2}}  \nonumber \\
&=&-(m^2/\hbar)\pi i\int d\vec{k'}\delta (k^{2}-k'^{2})|\phi _{\vec{k'}}^{(\mathrm{bg}%
+)}\rangle \langle \phi _{\vec{k'}}^{(\mathrm{bg}+)}|  \nonumber \\
&&+(m\hbar)\mathrm{P}\int d\vec{k'}\frac{|\phi
_{\vec{k'}}^{(\mathrm{bg}+)}\rangle \langle \phi
_{\vec{k'}}^{(\mathrm{bg}+)}|}{k^{2}-k'^{2}}.
\end{eqnarray}%
Here $\mathrm{P}\int $ refers to the principle value of the
integral.

Considering the relationship (\ref{fdelta}) between the scattering amplitude
$f_{1m_{z}}(k)$ and the phase shift $\delta _{1m_{z}}(k)$, it is easy to
prove that the phase shift $\delta _{1m_{z}}(k)$ corresponding to the
scattering amplitude (\ref{finalf}) is the one given in Eqs. (\ref{delta1a}%
) and (\ref{delta1}). It is pointed out that, this result can also
be proved with the method in Ref. (\cite{fano}).

\section{The scattering states with ingoing and outgoing boundary conditions}

In this appendix we prove the formula (\ref{nppsi}) in Appendix A.
We begin from the relationship \cite{julian} between the three
dimensional scattering states with ingoing and outgoing boundary
conditions:
\begin{eqnarray}
\langle \vec{r}|\phi _{\vec{k}}^{(\mathrm{bg}-)}\rangle =\langle \phi _{-%
\vec{k}}^{(\mathrm{bg}+)}|\vec{r}\rangle .  \label{c4}
\end{eqnarray}%
Considering the definitions (\ref{psik1mz}) and (\ref{psik1mzn}) of $|\psi
_{klm_{z}}^{(\mathrm{bg}\pm)}\rangle$, we can obtain
\begin{eqnarray}
\sum_{l,m_{z}}\langle \vec{r}|\psi _{klm_{z}}^{(\mathrm{bg}+)}\rangle
Y_{l}^{m_{z}}(\hat{k})^{\ast }=\sum_{l,m_{z}}\langle \psi _{klm_{z}}^{(%
\mathrm{bg}-)}|\vec{r}\rangle Y_{l}^{m_{z}}(-\hat{k})  \label{c5}
\end{eqnarray}%
We further define the one dimensional functions
$F_{k1}^{(\mathrm{bg}\pm
)}(r)$ as \cite%
{taylor}
\begin{eqnarray}
\langle \vec{r}|\psi _{k1m_{z}}^{(\mathrm{bg}\pm )}\rangle =i^{l}\frac{1}{\hbar}(\frac{m}{%
\pi k})^{\frac{1}{2}}\frac{1}{r}F_{k1}^{(\mathrm{bg}\pm
)}(r)Y_{1}^{m_{z}}(\hat{r}).  \label{c3}
\end{eqnarray}%
Using the relationships
\begin{eqnarray}
Y_{1}^{m_{z}}(-\hat{k}) &=&(-1)^{m_{z}+1}Y_{1}^{-m_{z}}(\hat{k})^{\ast };
\label{c6} \\
Y_{1}^{m_{z}}(\hat{r}) &=&(-1)^{m_{z}}Y_{1}^{-m_{z}}(\hat{r})^{\ast
}.
\end{eqnarray}%
we get%
\begin{eqnarray}
F_{k1}^{(\mathrm{bg}+)}(r)=F_{k1}^{(\mathrm{bg}-)}(r)^{\ast }.
\label{c8}
\end{eqnarray}

On the other hand, we know that $F_{k1}^{(\mathrm{bg}+)}(r)$ and $%
F_{k1}^{(\mathrm{bg}-)}(r)$ satisfy the same differential equation
\begin{eqnarray}
\left(
-\frac{d^{2}}{dr^{2}}+V^{(\mathrm{bg})}+\frac{2}{r^{2}}\right)
F_{k1}^{(\mathrm{bg}\pm )}(r)=k^{2}F_{k1}^{(\mathrm{bg}\pm )}(r).
\label{requation}
\end{eqnarray}%
with the same boundary condition $F_{k1}^{(\mathrm{bg}\pm )}(0)=0$.
Then $F_{k1}^{(\mathrm{bg}-)}(r)$ is proportional to $F_{k1}^{(%
\mathrm{bg}+)}(r)$. To calculate the ratio between $F_{k1}^{(\mathrm{bg}%
\pm )}(r)$, we consider their asymptotic behaviors in the limit
$r\rightarrow \infty $:
\begin{eqnarray}
F_{k1}^{(\mathrm{bg}+)}(r) &=&\hat{j}_{1}(kr)+kf_{k1}^{(\mathrm{bg}%
)}(k)\left[ \hat{n}_{1}(kr)+i\hat{j}_{1}(kr)\right];  \nonumber \\
\\
F_{k1}^{(\mathrm{bg}-)}(r) &=&\hat{j}_{1}(kr)+kf_{k1}^{(\mathrm{bg}%
)}(k)^{\ast }[\hat{n}_{1}(kr)-i\hat{j}_{1}(kr)]  \nonumber \\
\end{eqnarray}%
with $\hat{j}_{1}(x)$defined in (\ref{j1}) and
\begin{eqnarray}
\hat{n}_{1}(x)=-\frac{\cos x}{x}-\sin x
\end{eqnarray}
the irregular first order Riccati-Bessel function. Comparing the
coefficients
of $\hat{n}_{l}(kr)$, we have%
\begin{eqnarray}
F_{k1}^{(\mathrm{bg}-)}(r)=e^{-2i\delta _{1}^{(\mathrm{bg}%
)}(k)}F_{k1}^{(\mathrm{bg}-)}(r).  \label{c10}
\end{eqnarray}%
Here we have used%
\begin{eqnarray}
f_{k1}^{(\mathrm{bg})}(k) &=&-\frac{1}{ik-k\cot \delta _{1}^{(%
\mathrm{bg})}(k)}  \nonumber \\
&=&\frac{1}{k}\sin \delta _{1}^{(\mathrm{bg})}(k)e^{i\delta _{1}^{(%
\mathrm{bg})}(k)}.
\end{eqnarray}%
Substituting Eq. (\ref{c10}) into Eq. (\ref{c3}), we get Eq. (\ref{nppsi}).

\section{The expansion of the factor $g_{m_{z}}(k^{2})$}

In this appendix we prove the Eq. (\ref{gk2}). We firstly rewrite
the factor $g_{m_{z}}(k^{2})$ as
\begin{widetext}
\begin{eqnarray}
g_{m_{z}}(k^{2}) &=&\mathrm{Re}\langle \phi _{\mathrm{res}%
}^{(m_{z})}|WG_{+}^{(\mathrm{bg})}(0)W|\phi
_{\mathrm{res}}^{(m_{z})}\rangle
-k^{2}\mathrm{Re}\langle \phi _{\mathrm{res}}^{(m_{z})}|WG_{+}^{(\mathrm{bg}%
)}(0)G_{+}^{(\mathrm{bg})}(k^{2})W|\phi
_{\mathrm{res}}^{(m_{z})}\rangle . \label{expgk}
\end{eqnarray}
\end{widetext}
Here we have used the identity%
\begin{eqnarray}
G_{+}^{(\mathrm{{bg})}}(k^{2})=G_{+}^{(\mathrm{{bg})}}(0)-k^{2}G_{+}^{(%
\mathrm{{bg})}}(0)G_{+}^{(\mathrm{{bg})}}(k^{2}).
\end{eqnarray}%
The first term in the r.h.s. of (\ref{expgk}) is independent on $k$.
It contributes the constant term $g_{m_{z}}^{(0)}$ in Eq.
(\ref{gk2}).

On the other hand, the second term in Eq. (\ref{expgk}) can be
re-written as
\begin{eqnarray}
&&k^{2}\mathrm{Re}\langle \phi _{\mathrm{res}}^{(m_{z})}|WG_{+}^{(\mathrm{bg}%
)}(0)G_{+}^{(\mathrm{bg})}(k^{2})W|\phi
_{\mathrm{res}}^{(m_{z})}\rangle
\nonumber \\
&=&\frac{m^2k^{2}}{\hbar^{4}}\mathrm{Re}\lim_{\varsigma
_{1},\varsigma _{2}\rightarrow 0^{+}}\int
d\vec{k'}\frac{|\langle \phi _{\mathrm{res}}^{(m_{z})}|W|\phi _{\vec{k'}}^{(%
\mathrm{bg}+)}\rangle |^{2}}{(k'^{2}-i\varsigma
_{1})(k'^{2}-k^{2}-i\varsigma _{2})}\mathrm{.}\nonumber\\
\end{eqnarray}%
In the limit $k=0$, we have%
\begin{eqnarray}
&&\lim_{k\rightarrow 0}\mathrm{Re}\langle \phi _{\mathrm{res}%
}^{(m_{z})}|WG_{+}^{(\mathrm{bg})}(0)G_{+}^{(\mathrm{bg})}(k^{2})W|\phi _{%
\mathrm{res}}^{(m_{z})}\rangle   \nonumber \\
&\propto &\int dp\frac{|\langle \phi
_{\mathrm{res}}^{(m_{z})}|W|\psi _{p1m_{z}}^{(\mathrm{bg}+)}\rangle
|^{2}}{p^{3}}.  \label{c44}
\end{eqnarray}%
We know that the function $|\langle \phi
_{\mathrm{res}}^{(m_{z})}|W|\psi _{p1m_{z}}^{(\mathrm{bg}+)}\rangle
|^{2}$ decays to zero when $p\rightarrow
\infty $. On the other hand, as we have shown in Sec. III.B, the factor $|%
\mathscr{J}(k)|^{2}$ tends to a non-zero constant in the low energy
limit if the background scattering volume in the open channel is
finite. Using the
relationship (\ref{jost1}) between $\mathscr{J}(k)$, $\tilde{F}_{k1}^{(%
\mathrm{bg}\pm )}(r)$ and $|\psi _{p1m_{z}}^{(\mathrm{bg}+)}\rangle
$, and the low-energy behavior (\ref{ftuter}) of
$\tilde{F}_{k1}^{(\mathrm{bg}\pm
)}(r)$, it is easy to prove that the factor $|\langle \phi _{\mathrm{res}%
}^{(m_{z})}|W|\psi _{p1m_{z}}^{(\mathrm{bg}+)}\rangle |^{2}$ is
proportional to $p^{3}$ in the limit $p\rightarrow 0$. Therefore,
the above integration
in Eq. (\ref{c44}) converges to a finite constant in the limit $%
k^{2}\rightarrow 0$. Then the expansion in Eq. (\ref{gk2}) is proved
and we have
\begin{eqnarray}
g_{m_{z}}^{(2)}=-\langle \phi _{\mathrm{res}}^{(m_{z})}|WG_{+}^{(\mathrm{bg}%
)}(0)G_{+}^{(\mathrm{bg})}(0)W|\phi _{\mathrm{res}}^{(m_{z})}\rangle
\leq 0.
\end{eqnarray}

\section{The background Jost function}

In this appendix we calculate the Jost function $\mathscr{J}(k)$ of
the
background scattering state. To this end, we introduce a function $\bar{F}%
_{k1}^{(\mathrm{bg}\pm )}(r)=\tilde{F}_{k1}^{(\mathrm{bg}\pm
)}(r)/k^{2}$
where $\tilde{F}_{k1}^{(\mathrm{bg}\pm )}(r)$ is defined in Eq. (\ref{jost1}%
). It is apparent that $\bar{F}_{k1}^{(\mathrm{bg}\pm )}(r)$ is a
solution of the radial equation
\begin{eqnarray}
\left(
-\frac{d^{2}}{dr^{2}}+V^{(\mathrm{{{bg})}}}(r)+\frac{2}{r^{2}}\right)
\bar{F}_{k1}^{(\mathrm{bg})}(r)=k^{2}\bar{F}_{k1}^{(\mathrm{bg})}(r)\label{radical}
\end{eqnarray}%
with a $k$-independent boundary condition
\begin{eqnarray}
\bar{F}_{k1}^{(\mathrm{bg})}(r\rightarrow 0)\rightarrow r^{2}.
\end{eqnarray}

Following the spirit of quantum defect theory \cite{gao3}, we assume
the scattering potential $V^{(\mathrm{{{bg})}}}(r)$ can be
approximated as the van der Waals potential
$-\hbar^2\beta_{6}^4/(r^{6}m)$when $r$ is larger than a critical
distance $r_{0}$ which is much smaller than $\beta _{6}$. In the
region with $r<r_{0}$, $V^{(\mathrm{{{bg})}}}(r)$ is assumed to be
so large that the
atomic kinetic energy $k^{2}$ is negligible in comparing with $V^{(\mathrm{{{%
bg})}}}(r)$, and then $\bar{F}_{k1}^{(\mathrm{bg})}(r)$ is independent on $k$%
.

In the region $r>r_{0}$, $\bar{F}_{k1}^{(\mathrm{bg})}(r)$ is the
superposition of the two independent solutions $\chi _{\epsilon
1}^{(0)}(r)$ and $\kappa _{\epsilon 1}^{(0)}(r)$ of Eq.
(\ref{radical}) \cite{gao2} (in Ref. \cite{gao2}, $\chi _{\epsilon
1}^{(0)}(r)$ and $\kappa _{\epsilon 1}^{(0)}(r)$ are denoted as
$f_{\epsilon 1}^{(0)}(r)$ and $g_{\epsilon
1}^{(0)}(r)$):%
\begin{eqnarray}
\bar{F}_{k1}^{(\mathrm{bg})}(r)=\alpha _{k}\chi
_{k^{2}1}^{(0)}(r)+\beta _{k}\kappa _{k^{2}1}^{(0)}(r).
\end{eqnarray}%
In the short distance region with $r<<\beta _{6}$, $\chi
_{k^{2}1}^{(0)}(r)$ and $\kappa _{k^{2}1}^{(0)}(r)$ are almost
independent on $k$ \cite{gao3}.

The wave function $\bar{F}_{k1}^{(\mathrm{bg})}(r)$ in the two
regions are connected at point $r=r_{0}$, where
$\bar{F}_{k1}^{(\mathrm{bg})}(r)$, $\chi _{\epsilon 1}^{(0)}(r)$ and
$\kappa _{\epsilon 1}^{(0)}(r)$ are approximately independent on
$k$. Then we know that $\alpha _{k}$ and $\beta _{k}$ are
independent on $k$ and we have
\begin{eqnarray}
\bar{F}_{k1}^{(\mathrm{bg})}(r)=\alpha \chi
_{\epsilon_k1}^{(0)}(r)+\beta \kappa _{\epsilon_k1}^{(0)}(r).
\end{eqnarray}%
with $\epsilon_k=\hbar^2k^2/m$. The above equation yields%
\begin{eqnarray}
\tilde{F}_{k1}^{(\mathrm{bg})}(r)=k^{2}\alpha \chi
_{\epsilon_k1}^{(0)}(r)+k^{2}\beta \kappa
_{\epsilon_k1}^{(0)}(r).\label{ss}
\end{eqnarray}

In the region $r\rightarrow \infty $, the asymptotic behaviors of
$\chi _{\epsilon_k1}^{(0)}(r)$ and $\kappa _{\epsilon_k1}^{(0)}(r)$
can be expressed in terms of $Z_{ij}(k)$ ($i,j=f,g$) defined in
\cite{gao2}
\begin{eqnarray}
\chi _{\epsilon_k1}^{(0)}(r) &\rightarrow &\sqrt{\frac{2}{\pi
k}}\left[
Z_{ff}\sin \left( kr-\frac{\pi }{2}\right) -Z_{fg}\cos \left( kr-\frac{\pi }{%
2}\right) \right] ;  \nonumber \\
&&  \label{ff1} \\
\kappa _{\epsilon_k1}^{(0)}(r) &\rightarrow &\sqrt{\frac{2}{\pi
k}}\left[
Z_{gf}\sin \left( kr-\frac{\pi }{2}\right) -Z_{gg}\cos \left( kr-\frac{\pi }{%
2}\right) \right] .  \nonumber \\
&&  \label{g1}
\end{eqnarray}%
On the other hand we know that in the same limit we have
\cite{taylor}
\begin{eqnarray}
\tilde{F}_{k1}^{(\mathrm{bg})}(r) &=&\mathscr{J}(k)\left[ e^{i\delta _{1}^{(%
\mathrm{bg})}(k)}\sin \left( kr-\frac{\pi }{2}+\delta _{1}^{(\mathrm{bg}%
)}(k)\right) \right] .  \nonumber  \label{sc} \\
&&
\end{eqnarray}%
Together with Eqs. (\ref{ss}), (\ref{ff1}), (\ref{g1}) as well as (\ref{sc}%
), we obtain the expression of (\ref{jost}) of
$|\mathscr{J}(k)|^{2}$ and the background phase shift:
\begin{eqnarray}
\tan \delta _{1}^{(\mathrm{bg})}=-\frac{K_{l=1}^{0}Z_{gg}-Z_{fg}}{%
K_{l=1}^{0}Z_{gf}-Z_{ff}}.\label{tgdeltabg}
\end{eqnarray}%
where $D_{ij}(k)=(k\beta _{6})^{3/2}Z_{ij}(k)$ and
$K_{l=1}^{0}=-\beta /\alpha $ is the one in Sec. III. The result
(\ref{tgdeltabg}) is also given in \cite{gao1}.

\end{subappendices}

\end{document}